\documentclass[reprint,preprintnumbers,nofootinbib,superscriptaddress,amsmath,amssymb,aps,prd]{revtex4-2}
\usepackage{graphicx,hyperref,braket,booktabs,orcidlink}
\graphicspath{{img/}}

\renewcommand{\vec}[1]{\mathbf{#1}}

\newcommand{\md}{\mathrm{d}}

\newcommand{\lmax}{l_{\mathrm{max}}}

\DeclareMathOperator{\Tr}{Tr}

\hypersetup{
	unicode=true,
	pdftoolbar=true,
	pdfmenubar=true,
	pdffitwindow=false,
	pdfstartview={FitH},
	pdftitle={Towards a Quantum Simulation of Nonlinear Sigma Models with a Topological Term},
	pdfauthor={J.Y.~Araz, S.~Schenk, M.~Spannowsky}, 
	pdfkeywords={Quantum Simulations, Tensor Networks, Nonlinear Sigma Model},
	pdfnewwindow=true,
	colorlinks=true,
	linkcolor=blue,
	citecolor=magenta,
	filecolor=magenta,
	urlcolor=blue
}

\begin{document}

\title{Towards a Quantum Simulation of Nonlinear Sigma Models with a Topological Term}

\author{Jack Y.~Araz\orcidlink{0000-0001-8721-8042}}
\email{jack.araz@durham.ac.uk}
\affiliation{Institute for Particle Physics Phenomenology, Durham University, South Road, Durham DH1 3LE, United Kingdom}

\author{Sebastian Schenk\orcidlink{0000-0002-9143-8870}}
\email{schenkse@uni-mainz.de}
\affiliation{Institute for Particle Physics Phenomenology, Durham University, South Road, Durham DH1 3LE, United Kingdom}
\affiliation{PRISMA$^{+}$ Cluster of Excellence \& Mainz Institute for Theoretical Physics, Johannes Gutenberg-Universit\"at
Mainz, 55099 Mainz, Germany}

\author{Michael Spannowsky\orcidlink{0000-0002-8362-0576}}
\email{michael.spannowsky@durham.ac.uk}
\affiliation{Institute for Particle Physics Phenomenology, Durham University, South Road, Durham DH1 3LE, United Kingdom}

\preprint{IPPP/22/68}
\preprint{MITP-22-079}

\begin{abstract}
We determine the mass gap of a two-dimensional $O(3)$ nonlinear sigma model augmented with a topological $\theta$-term using tensor network and digital quantum algorithms.
As proof of principle, we consider the example $\theta = \pi$ and study its critical behaviour on a quantum simulator by examining the entanglement entropy of the ground state.
We confirm that the quantum theory is massless in the strong-coupling regime, in agreement with analytical results.
However, we also highlight the limitations of current quantum algorithms, designed for noisy intermediate-scale quantum devices, in the theory simulation at weak coupling.
Finally, we compare the performance of our quantum algorithms to classical tensor network methods.
\end{abstract}

\maketitle

\section{Introduction}
\label{sec:introduction}

Nonlinear sigma models have often been a test bed for exploring the intricate relationship between high-energy physics and condensed matter systems.
In particular, the $O(3)$ nonlinear sigma model in two dimensions exhibits various non-trivial features of quantum field theories (QFTs).
These include asymptotic freedom and a dynamical generation of a strong scale~\cite{Polyakov:1975rr}, or the emergence of instantons and merons~\cite{Polyakov:1975yp, Gross:1977wu}.
It furthermore shares many of these features with four-dimensional Yang-Mills theory~\cite{Polyakov:1977vm}.

In two dimensions, the $O(3)$ nonlinear sigma model allows for non-trivial $\theta$-vacua.
Their presence has two critical implications concerning the structure of the underlying QFT.
On the one hand, physically, they can dramatically change the fundamental properties of the quantum theory.
In general, the latter exhibits different features within different topological sectors.
For instance, the nonlinear sigma model with a topological term of $\theta = 2\pi S$ is closely related to a quantum spin chain with spin $S$~\cite{Haldane:1983ru}.
This observation can be exploited to obtain the mass gap of the quantum theory in different topological sectors (see, e.g.,~\cite{Haldane:1982rj}).
For instance, it is well known that the nonlinear sigma model without a topological term, $\theta = 0$, is gapped~\cite{Shigemitsu:1980tx, Gliozzi:1985jn, Hasenfratz:1990zz, Hasenfratz:1990ab,sachdev2011quantum}.
In strong contrast, the same theory with $\theta = \pi$ is massless at the quantum level for any value of the coupling constant~\cite{Shankar:1989ee}.
In this scenario, a conformal field theory (CFT) describes the underlying quantum theory's critical behaviour by the vanishing mass gap.
In the strong-coupling regime of the nonlinear sigma model, the latter is equivalent to the Wess-Zumino-Novikov-Witten CFT, with central charge $c=1$~\cite{Wess:1971yu, Novikov:1982ei, Witten:1983tw, Witten:1983ar, Affleck:1987ch, Affleck:1988nt, Shankar:1989ee}.
Similarly, the central charge of the CFT should approach $c=2$ in the weak-coupling limit~\cite{Shankar:1989ee}.

On the other hand, in practice, the computational treatment of the QFT is altered.
In particular, a numerical investigation is drastically hampered due to the so-called sign problem.
Any $\theta$-term that renders the action imaginary will turn the path integral into a highly oscillatory object.
A precise evaluation of the latter will typically fail due to numerical cancellations between contributions of different complex phases.
In fact, this is the source of the sign problem and is one of the significant drawbacks in applying Monte Carlo techniques to non-trivial QFTs, such as quantum chromodynamics on the lattice (see, e.g.,~\cite{deForcrand:2009zkb}).
However, in some special situations, these problems can be bypassed to a certain degree, notably in nonlinear sigma models~\cite{Bietenholz:1995ks,Bogli:2011aa,deForcrand:2012se,Azcoiti:2007cg,Alles:2007br,Azcoiti:2012ws}.\footnote{For a recent assessment of $\theta$-vacua and their lattice regularization in these models, see~\cite{Nguyen:2022aaq}.}

Tensor Network (TN) methods have been designed to overcome the sign problem.
They have successfully addressed various lattice gauge theory questions~\cite{Banuls:2018jag,Banuls:2019rao,Dalmonte:2016alw,Magnifico:2020bqt} and, in particular, also nonlinear sigma models at $\theta=0$~\cite{Bruckmann:2018usp} and $\theta=\pi$~\cite{Tang:2021uge}.
Although TNs constitute a promising approach to studying QFTs on the lattice, they also suffer from several shortcomings.
For instance, it tensors of large dimensions\footnote{Technically speaking, the bond dimensions of the TN are often large, $\mathcal{O}(\geq 10^3)$.} are typically required to approximate the underlying quantum theory sufficiently.
Additionally, despite the availability of various contraction algorithms, simulating TNs for lattice QFTs beyond two-dimensional spacetimes becomes computationally highly challenging.
This is mainly due to the attempt to solve a quantum system with a classical approach where the quantum nature has been compensated by extensive resources.
More precisely, the structure of entanglement is captured with large auxiliary dimensions.
Quantum computing (QC) techniques may allow us to overcome this issue by eliminating the need for extensive resources by intrinsically retaining the quantum nature of the problem.

The simulation of QFTs on quantum devices has been discussed in pioneering studies~\cite{Jordan:2012xnu, Jordan:2011ci} in which various quantum algorithms have been proposed, such as quantum Monte Carlo, Hamiltonian simulation, and imaginary-time evolution algorithms.
The Hamiltonian simulation, similar to TN approaches, allows quantum algorithms to avoid the sign problem mentioned above.
However, it is essential to note that these techniques have their own limitations.
Due to the limited number of ``noise free" qubits or qubits to control the error, the embedding of the Hamiltonian is significantly restricted.
For scenarios with infinite Hilbert space dimensions, Hamiltonian truncation is especially crucial~\cite{PhysRevA.105.052405, Tong2022provablyaccurate}.

In particular, nonlinear sigma models have been investigated in spin-lattice systems~\cite{Schenk:2020lea} and with quantum computing techniques~\cite{Alexandru:2019ozf, Alexandru:2022son}.
In the latter case, it is demonstrated that nonlinear sigma models can have a qubit-efficient description through a fuzzy-sphere representation.
Although this allows for a Hamiltonian truncation that is suitable for quantum time evolution algorithms, it is not prone to a straightforward generalisation to nonlinear sigma models that feature a richer structure.
In this work, we go beyond this limitation and augment a two-dimensional nonlinear sigma model with a topological $\theta$-term.
In particular, using quantum-gate simulators, we aim to study the entanglement entropy of the vacuum to investigate the critical behaviour and determine the mass gap of the quantum theory.

In addition to the simulation of QFTs, both TN-inspired quantum circuits and conventional TN methods have also been used in various machine-learning applications.
It has been shown that their relation with quantum many-body systems can be used to achieve more interpretable networks~\cite{Huggins_2019, Araz:2022haf, Araz:2021zwu, PhysRevA.101.010301}.

This work is organised as follows.
In Section~\ref{sec:hamiltonian}, we review the Hamiltonian formulation of the $O(3)$ nonlinear sigma model at $\theta=\pi$ in terms of angular momentum variables on a one-dimensional spin chain.
Section~\ref{sec:quantumsimulation} shows how these can be embedded on a quantum computer.
In particular, we use this approach to compute the bipartite entanglement entropy associated with the half chain and we confirm the vanishing mass gap of the theory in Section~\ref{sec:results}.
Finally, we briefly summarise our results and conclude in Section~\ref{sec:conclusion}.

\section{Hamiltonian formulation of nonlinear sigma models}
\label{sec:hamiltonian}

The field content of a general $O(3)$ nonlinear sigma model is given by a real vector field $\vec{n}$ that takes values on a sphere, $S^2$, i.e.~it is normalised to $\vec{n}^2 = 1$.
In a two-dimensional Euclidean spacetime, the associated action is commonly written as
\begin{equation}
	S = \frac{1}{2g^2} \int \md^2 x \, \left( \partial_{\mu} \vec{n} \right)^2 \, .
\end{equation}
Here, $g$ is the dimensionless coupling constant, and we consider Euclidean coordinates $\tau$ and $x$.
At the classical level, the vector field is massless.
This remains true to all orders in perturbation theory.
Nevertheless, it can be shown that the quantum theory is gapped~\cite{Shigemitsu:1980tx, Gliozzi:1985jn, Hasenfratz:1990zz, Hasenfratz:1990ab,sachdev2011quantum}.

As the two-dimensional $O(3)$ nonlinear sigma model admits instanton solutions (see, e.g.,~\cite{Polyakov:1975yp, Gross:1977wu}), it is feasible to consider an additional topological term in this theory.
Along these lines, we can distinguish finite-action field configurations by their topological charge, $Q = \int \md^2 x \, \rho_Q$, where
\begin{equation}
	\rho_Q = \frac{1}{4\pi} \epsilon_{abc} n^a \partial_x n^b \partial_\tau n^c
\end{equation}
is the topological charge density.
These field configurations, in turn, contribute a finite $\theta$-term to the action,
\begin{equation}
	S_{\theta} = \frac{1}{2g^2} \int \md^2 x \, \left( \partial_{\mu} \vec{n} \right)^2 + i \theta Q \, .
\end{equation}
Since the topological charge is an integer, the action is $2\pi$-periodic with respect to $\theta$.
In the following, we aim to investigate the topological term's effect on the quantum theory's mass gap.
For concreteness, let us focus on the case $\theta = \pi$ in the following.
As we will see momentarily, this choice allows for a simple Hamiltonian formulation of the QFT.
We will comment on the general case later in this work.

For a numerical investigation of the two-dimensional $O(3)$ nonlinear sigma model using quantum algorithms, a suitable Hamiltonian formulation of the former is needed~\cite{Hamer:1978uq, Hamer:1978ew, Shigemitsu:1980tx}.
Somewhat fortunately, this allows us to treat the two-dimensional theory from an effectively one-dimensional perspective as follows.
First, one considers the theory on a discrete spatial axis while keeping the time coordinate continuous at the same time.
In this scenario, the time derivative of the kinetic term can be identified with an angular momentum per lattice site.
Therefore, a one-dimensional chain of coupled quantum rotors can describe the two-dimensional field theory.
In particular, for the nonlinear sigma model at $\theta = \pi$, the Hamiltonian can be written as~\cite{Shankar:1989ee}
\begin{equation}
	H = \frac{1}{2\beta} \sum_{k=1}^N \vec{L}_k^2 + \beta \sum_{k=1}^{N-1} \vec{n}_{k} \vec{n}_{k+1} \, .
\label{eq:Hamiltonian}
\end{equation}
Here, we set the lattice spacing in the spatial direction to $a=1$ and make the replacement $\beta = 1/g^2$.
In this setup, $\vec{L}_k$ denotes a (modified) quantum mechanical angular momentum operator acting on the $k$-th site of the spin chain.\footnote{We again remark that this spin chain fully characterises the two-dimensional nonlinear sigma model. This is due to the choice of Hamiltonian variables (see, e.g.,~\cite{Hamer:1978ew}).}
While this operator acts on the local Hilbert space at each site, the second term of the Hamiltonian corresponds to the interactions of neighbouring sites, which we will characterise momentarily.
Note that, at this stage, we impose open boundary conditions to keep the notation simple.
This is why the summation of the second term is terminated at the $N$-th site.
In practice, we will later use periodic boundary conditions in the simulation.

Before we continue, let us make a few essential remarks on the Hamiltonian formulation~\eqref{eq:Hamiltonian}, closely following~\cite{Shankar:1989ee}.
Naively, the angular momentum operator represents a particle moving on a unit sphere with coordinate $\vec{n}$.
However, it turns out that, in our scenario, the topological $\theta$-term for $\theta = \pi$ corresponds to a vector potential $\vec{A} (\vec{n})$.
Physically, the latter is sourced by a magnetic monopole located at the centre of the unit sphere.
That is, the particle on the sphere is moving in the monopole potential.
The angular momentum operator in the position space representation is modified accordingly, such that it takes the form $\vec{L} = \vec{n} \times \left( - i \nabla - \vec{A} \right) - \vec{n}$.
Therefore, finally, we can systematically construct a suitable basis for the local Hilbert space at each site of the spin chain using monopole harmonics.
At the same time, this clearly prevents a straightforward generalisation of this method to arbitrary $\theta$ (as the Hamiltonian formulation is closely related to the quantised monopole background).

\subsection{Constructing the local Hilbert space}

Within our approach, we are interested in the low-lying excitations in the spectrum of the Hamiltonian~\eqref{eq:Hamiltonian}.
That means, quantum mechanically, we first need to determine the local Hilbert space associated with each site of the spin chain.
As the angular momentum operator is the generator of rotations, it is intuitive to use an eigenbasis of the former.
It therefore acts on the local Hilbert space at the $k$-th site as
\begin{equation}
\begin{split}
	\vec{L}_k^2 \ket{qlm}_k &= l(l+1) \ket{qlm}_k \, , \\
	L_k^{z} \ket{qlm}_k &= m \ket{qlm}_k \, ,
\end{split}
\label{eq:O3_ladder}
\end{equation}
where $L_k^z$ denotes the $z$-component of $\vec{L}_k$.
Following our earlier discussion, we have introduced the monopole charge $q$, which, in our example, takes the value $q = 1/2$.\footnote{Note that the scenario $q=0$ reduces to the well-known quantum mechanical angular momentum basis.}
In this case, $l$ is a positive half-integer, $l = 1/2,\ 3/2,\ \ldots$, and the projection quantum number takes values $m = -l,\ \ldots,\ l$.
Similarly, a discussion on how the operators $\vec{n}_k$ act on the local Hilbert space at each site can be found in Appendix~\ref{app:matrix_elements}.
For more details on this basis, we refer the reader to~\cite{Wu:1976ge, Wu:1977qk}.

Finally, the multiparticle state of the spin chain, in this basis, is then schematically characterised by the tensor product
\begin{equation}
	\ket{\psi} = \bigotimes_{k=1}^N \alpha_k \ket{qlm}_k \, ,
\end{equation}
with coefficients $\alpha_k$.
In this basis, we characterise all necessary matrix elements of the operators belonging to the Hamiltonian $H$ in Appendix~\ref{app:matrix_elements}.

\subsection{Truncating the local Hilbert space}

It is evident from Eq.~\eqref{eq:O3_ladder} that the spectrum of the angular momentum operator $\vec{L}^2$ is not bounded from above.
Therefore, the local Hilbert space associated with each site is infinite-dimensional, as we expect from a generic QFT perspective.
Our approach only applies to finite vector spaces, so we must use a suitable truncation for each local Hilbert space.
In practice, this means that we only consider quantum states up to a specific (maximal) orbital angular momentum quantum number $\lmax$.
For simplicity, we choose the same truncation for the local Hilbert space at each site.
Each Hilbert space is then of dimension
\begin{equation}
	\dim \mathcal{H} = \sum_{l=1/2}^{\lmax} \left(2l+1\right) = \lmax (\lmax + 2) + \frac{3}{4} \, .
\end{equation}
In principle, the truncation $\lmax$ is a free parameter of our approach, which has to be treated with care as it may neglect significant parts of the Hilbert space if chosen too small.
For instance, crucially, the truncation violates the nonlinear constraint on the vector field, $\vec{n}^2 = 1$.
Nevertheless, we expect these complications to be negligible for sufficiently large $\lmax$, which has to be carefully checked throughout the simulation.
For a more detailed discussion of this, see~\cite{Tang:2021uge} (and for the scenario $\theta = 0$ also \cite{Bruckmann:2018usp}).

In principle, for the case of nonlinear sigma models, one could also impose a different truncation scheme of the local Hilbert spaces.
One particular example is based on the fuzzy sphere, inspired by noncommutative geometry~\cite{Madore:1991bw,Alexandru:2019ozf}.
Here, the space of continuous function of $\vec{n}$ is replaced by the finite-dimensional algebra of generators of $SU(2)$, $n_i \to J_i$.
As such, it is not straightforward to generalise this representation to include a topological $\theta$-term.
Luckily, it will turn out that the smallest truncation that we can simulate on a quantum device is easily constructed in our truncation scheme.
A recent systematic, detailed discussion of the truncation of bosonic QFTs can be found in~\cite{Alexandru:2022son}.

\section{Quantum simulation}
\label{sec:quantumsimulation}

In the following, we briefly highlight our computational methods to investigate the two-dimensional nonlinear sigma model featuring a topological $\theta$-term.
We focus on the implementation of a suitable quantum circuit.
Furthermore, in practice, throughout the rest of this work, we impose periodic boundary conditions.

\subsection{Implementation of the quantum circuit}
\label{sec:vqe}

A Variational Quantum Eigensolver (VQE)~\cite{vqe2014, Tilly:2021jem,Cerezo:2020jpv} allows for a flexible ansatz to determine the ground or excited states of a given Hamiltonian.
It uses a quantum device to store a parametrised wave function and optimises parameters for the energy expectation value via external, classical minimisation techniques.
The parametrisation of the VQE is achieved by applying a series of unitary gates to the initial state, $U^\dagger(\phi)\ket{0}$.
In this case, the desired Hamiltonian expectation value is given by $\lambda \simeq \braket{0 | U(\phi)\ H\ U^\dagger(\phi) | 0}$.
Here, $\phi$ collectively denotes the trainable parameters of the VQE, and $\lambda$ stands for the true expectation value of the given Hamiltonian.

\begin{figure*}[t]
    \centering
    \includegraphics[scale=0.4275]{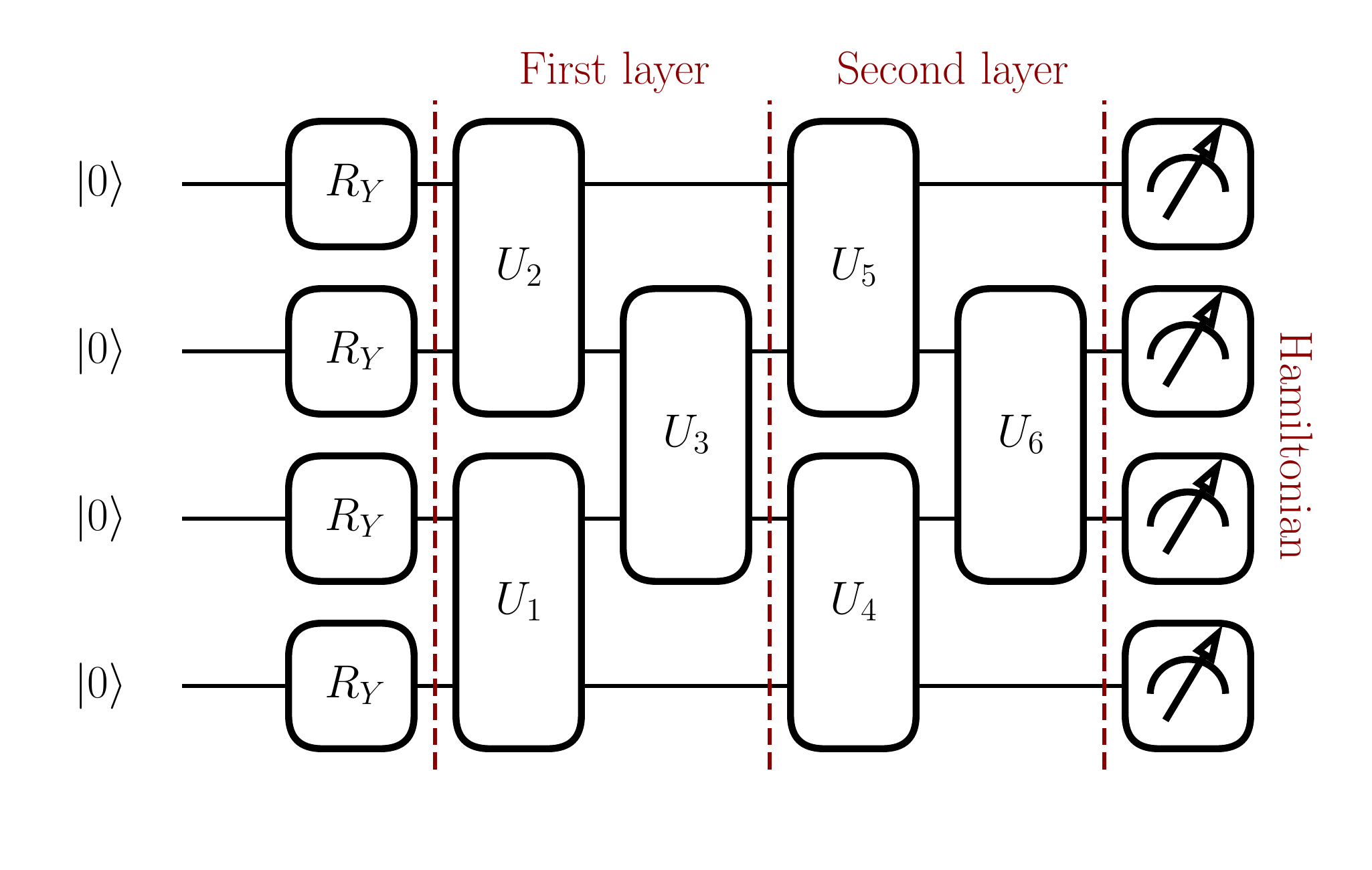}
    \includegraphics[scale=0.4275]{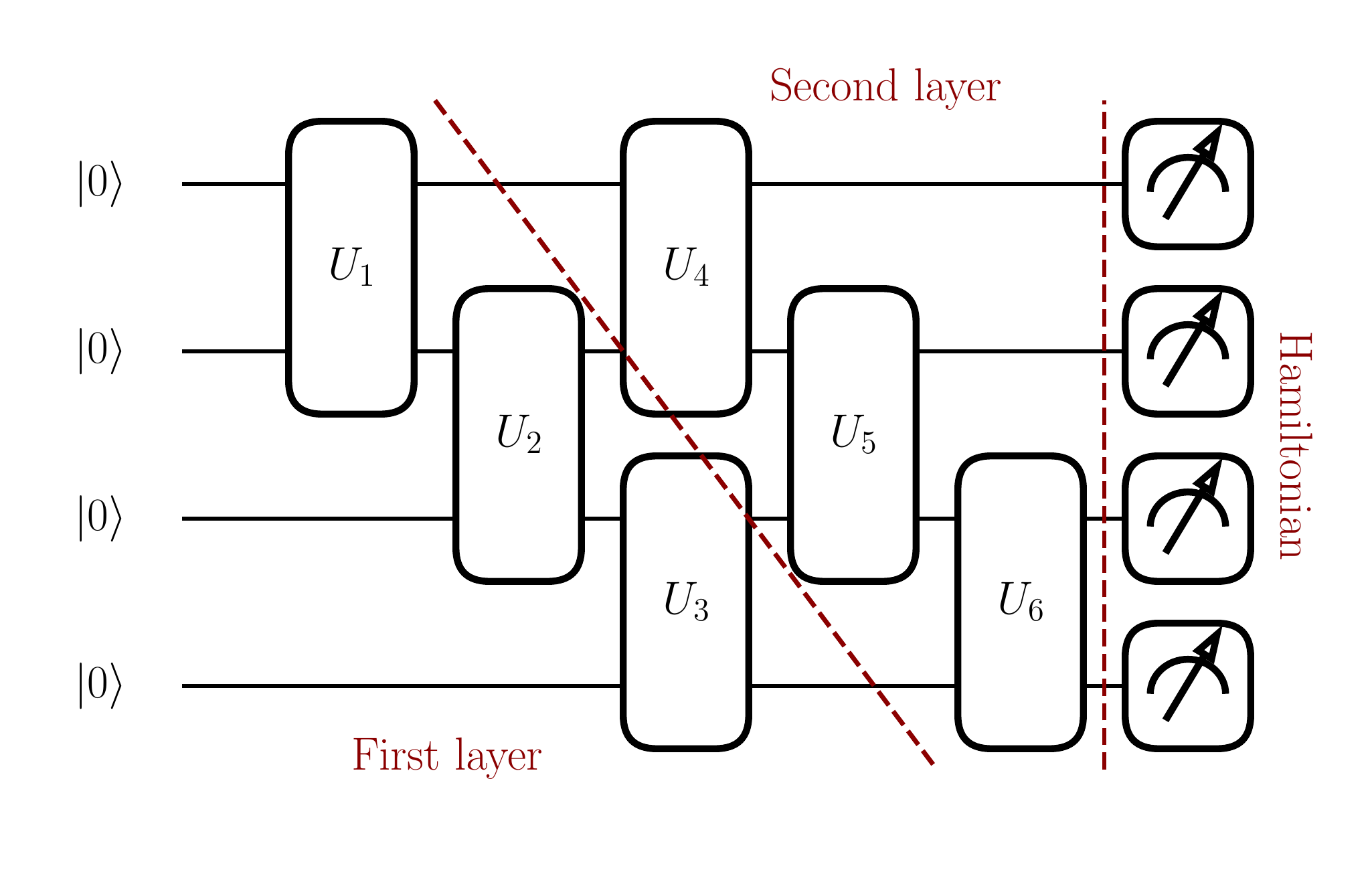}
    \caption{Quantum circuit architectures implementing a simplified two-design (left) and an MPS-inspired structure (right). Here, $R_Y$ illustrates the rotation gate around the Pauli-$Y$ axis, and $U_i$ represents the two-qubit unitary operator. Each circuit involves two layers acting on the Hamiltonian.}
    \label{fig:vqe}
\end{figure*}

The operator $U(\phi)$ generally embeds a series of rotation and entangling gates to diagonalise the given Hamiltonian.
However, the given ansatz is crucial to achieve the necessary accuracy for the ground state estimation.
In this work, we employ two types of $U(\phi)$ architectures, namely a so-called simplified two-design~\cite{Cerezo2021} as well as Matrix Product States (MPS) inspired quantum circuits~\cite{Huggins_2019}.
Both are shown in Fig.~\ref{fig:vqe}.

The simplified two-design embeds a multiscale renormalisation ansatz (MERA)~\cite{Vidal:2008zz} like entanglement structure to the circuit, thereby letting the ansatz capture complex correlations between qubits.
The left panel of Fig.~\ref{fig:vqe} illustrates a two-layer example of this scenario.
Regardless of the number of layers, this ansatz initialises each qubit with a rotation around the Pauli-$Y$ axis with a rotation angle that is randomly initialised and updated by the optimisation algorithm.
In the following, two-qubit entanglement is captured by applying unitary gates in a MERA-like pattern.
Each unitary gate applies a controlled-Z gate to the input qubits before each qubit is then rotated around the Pauli-$Y$ axis.

The MPS-inspired quantum circuit follows a staircase-like architecture, where the additional bond dimensions are captured by adding extra layers instead of introducing auxiliary qubits~\cite{Rudolph:2022dml}.
For example, the right panel of Fig.~\ref{fig:vqe} illustrates a two-layer MPS structure involving four qubits.
Each two-qubit unitary gate applies two $R_Y(\phi_i)$ rotations to each qubit, followed by a controlled-X gate (or, in other words, a CNOT gate).

Contrary to classical computing methods, a quantum circuit cannot be written out of any given tensors.
Hence, any required information has to be embedded into the circuit via the device's set of available gate operations.
In the case of the Hamiltonian embedding, any given set of operations has to be mapped on a set of Pauli gates to be embedded on the circuit, $H \supset \{ I, \sigma_X, \sigma_Y, \sigma_Z \}$.
Consequently, the Hamiltonian is written as a combination of Kronecker products involving Pauli matrices.
In practice, for any given truncation scheme of the Hamiltonian, this requires each of its terms to have a $2^d \times 2^d$ shape, i.e.~the local Hilbert space at each site has to be $2^d$-dimensional.
In practice, we use a truncation parameter of $\lmax = 1/2$ to be able to simulate as many lattice sites as possible.
This allows each lattice site to be represented by a single qubit.
It is highly challenging to go beyond this truncation parameter for any practical purpose.
We will comment on this limitation later in this work.
A suitable decomposition of the Hamiltonian~\eqref{eq:Hamiltonian} in terms of the above quantum circuits is briefly presented in Appendix~\ref{app:hamiltonian_embedding}.

We use various layers for different lattice sites to reach the necessary accuracy for the ground state energy and its associated entanglement entropy.
The number of layers and parameters for each architecture is presented in Tab.~\ref{tab:opt_num}.
Each circuit has been trained via an \textsc{Adam} optimisation algorithm~\cite{Kingma:2014vow} with an 0.1 initial learning rate for 2000 epochs\footnote{It is essential to note that we also employ a quantum natural gradient descent~\cite{Stokes_2020}. However, we do not find a significant improvement. In fact, due to a large number of parameters, the algorithm is significantly slower.}.
The learning rate is then halved every 200 epochs if the expectation value stops reducing in the last 20 epochs.
In addition, the optimisation is stopped if the precision of the expectation value cannot improve above $10^{-6}$ for more than 50 epochs.
For the simulation of the quantum circuits, we employ the \textsc{PennyLane} package~\cite{Bergholm:2018cyq}, which provides gradients of the expectation value for trainable parameters via the parameter-shift method~\cite{Mitarai_2018, Schuld_2019}.
The overall error estimation is achieved by reinitialising the circuit, optimising it 100 independent times, and taking the standard deviation of the outcome.
Hence, the error estimation for the quantum circuit algorithm is not due to the quantum device but to the training/optimisation algorithm.
That is, we assume ideal conditions for the quantum simulation.

\begin{table}[t]
	\centering
    	\begin{tabular}{c|cc|cc}
    		\toprule
    		& \multicolumn{2}{c|}{Simplified two-design} & \multicolumn{2}{c}{MPS} \\
		\midrule
		Sites & Layers & Parameters & Layers & Parameters \\
		\midrule
		4 & 5 & 34 & 3 & 18 \\
		6 & 10 & 106 & 8 & 80 \\
		8 & 39 & 554 & 21 & 294 \\
		10 & 50 & 910 & 28 & 504 \\
		\bottomrule
	\end{tabular}
    \caption{Number of layers as well as parameters used in the quantum circuits of the respective architecture. These are shown as a function of the number of lattice sites.}
    \label{tab:opt_num}
\end{table}

\begin{table}[t]
	\centering
    	\begin{tabular}{c|cc|cc}
    		\toprule
    		& \multicolumn{2}{c|}{$\lmax=1/2$} & \multicolumn{2}{c}{$\lmax=3/2$} \\
		\midrule
		Sites & Bond dim. & Parameters & Bond dim. & Parameters \\
		\midrule
		4 & 4 & 128 & 36 & 31104 \\
		6 & 8 & 768 & 207 & 1542564 \\
		8 & 16 & 4096 & 564 & 15268608 \\
		10 & 32 & 20480 & 903 & 48924540 \\
		\bottomrule
	\end{tabular}
    \caption{Maximum bond dimension and estimated number of parameters used in the DMRG approach, involving MPSs for different Hilbert space truncations. These are shown as a function of the number of lattice sites.}
    \label{tab:opt_num_tn}
\end{table}

\subsection{Comparison to a tensor network approach}
\label{subsec:tensornetworks}

As a nontrivial crosscheck, we also compare our results to methods involving classical TN implementations of MPS.
In this framework, a recent study of the nonlinear sigma model with a topological term of $\theta = \pi$ has been put forward in~\cite{Tang:2021uge}.

In particular, we use a Density Matrix Renormalization Group (DMRG) ansatz~\cite{White:1992zz,White:1993zza,Schollwock:2005zz,schollwock2011density} to find the ground state of the quantum theory.
The general strategy of DMRG is to find the dominant eigenvectors of a large matrix (in our case, $H$) by iterating over each matrix at each lattice site individually while keeping the others fixed.
Naively, the algorithm computes the dominant eigenvalues of the matrix associated with each site from the effective Hamiltonian corresponding to the interactions with the environment.
In a working scenario, we finally end up with an MPS corresponding to the system's ground state.
In practice, we use the \textsc{ITensor} library~\cite{Fishman:2020gel} to implement this method.
The overall error estimation is then given by the truncation error cutoff of the DMRG procedure.

In contrast to the digital quantum algorithms presented previously, the DMRG ansatz involving MPSs allows us to go beyond the truncation limitation of $\lmax = 1/2$, as the method itself is independent of the local Hilbert space dimensions.
An estimated number of required bond dimensions and parameters of the MPS is illustrated in Tab.~\ref{tab:opt_num_tn}, which we will comment on later.

\section{Results}
\label{sec:results}

In the following, we collect the results of the quantum simulation of the two-dimensional nonlinear sigma model featuring a topological term at $\theta = \pi$.
In particular, by the methods presented in the previous section, we first construct the ground state of the Hamiltonian and then study its entanglement entropy.
This will enable us to study the critical behaviour and determine the mass gap of the quantum theory.

\subsection{The ground-state energy density}
\label{subsec:groundstate}

\begin{figure*}[t]
	\centering
	\includegraphics[width=0.75\columnwidth]{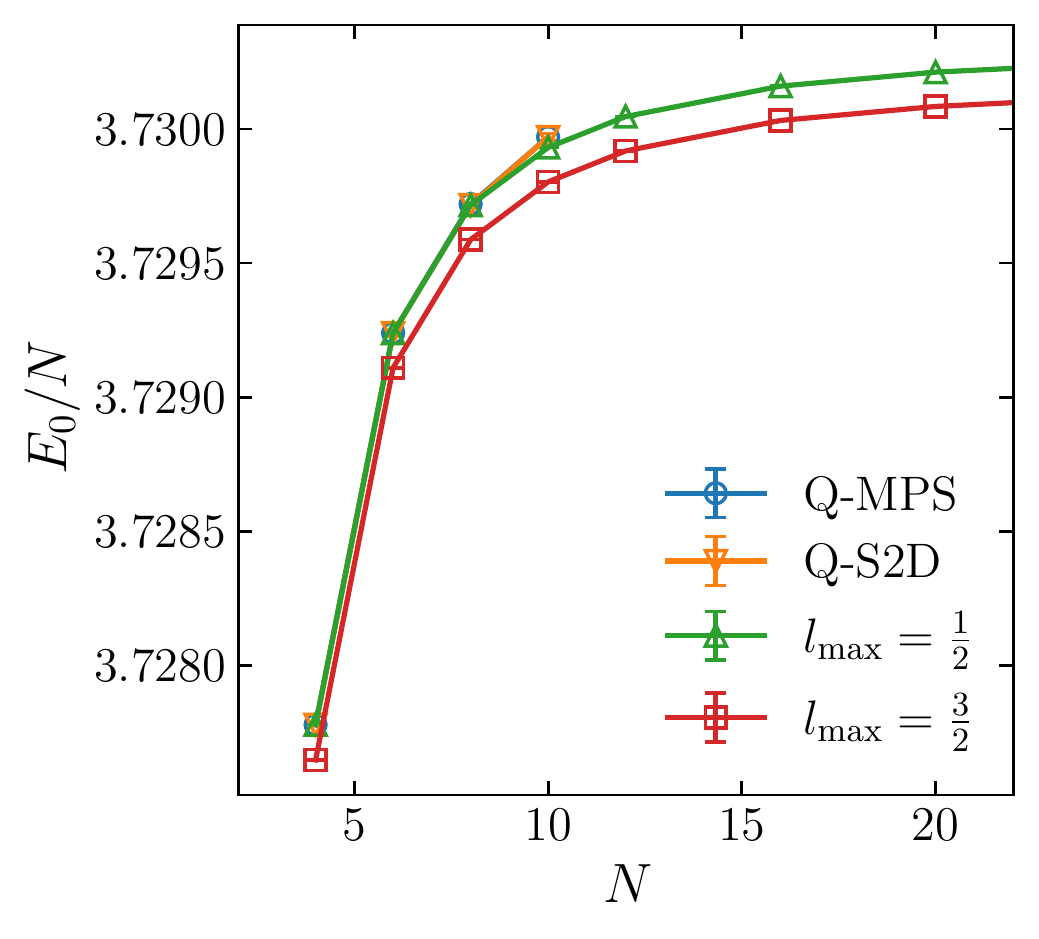}
	\hspace{2.75cm}
	\includegraphics[width=0.71\columnwidth]{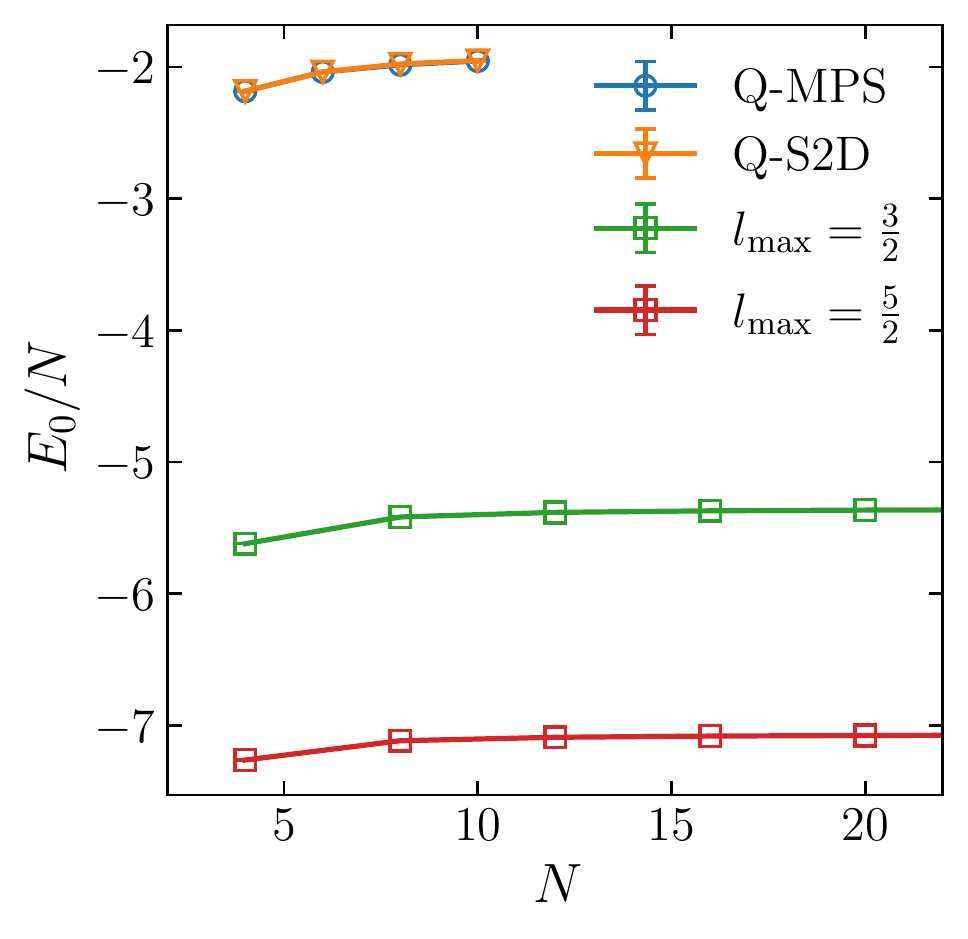}
	\caption{Ground-state energy density as a function of the number of lattice sites in the strong-coupling regime (left) and the weak-coupling regime (right). We fix the coupling constant to $\beta = 1/10$ and $\beta = 10$, respectively. The blue and orange lines denote the simplified two-design and the MPS-inspired quantum circuits, while the green and red lines correspond to conventional TNs, characterised by the truncation parameter $\lmax$. In both scenarios, we impose periodic boundary conditions. The error bars illustrate the uncertainties due to the optimisation algorithm (for the quantum circuits) and the DMRG truncation error cutoff (for the TNs), respectively. Carefully note that both drastically underestimate the true error done by the Hamiltonian truncation procedure (see main text).}
\label{fig:groundstate_O3_theta}
\end{figure*}

In the first step, we aim to determine the ground state of the quantum theory.
In practice, this corresponds to finding the lowest excitation in the spectrum of the Hamiltonian~\eqref{eq:Hamiltonian}.
As benchmark examples, we consider the strong-coupling limit, $\beta \to 0$, and the weak-coupling limit, $\beta \to \infty$ (note here that both refer to the inverse of the nonlinear sigma model coupling constant $g^2$).
We illustrate the respective ground-state energy densities in both regimes as a function of the system size in Fig.~\ref{fig:groundstate_O3_theta}.
However, we remark that we could also choose any other intermediate value of the coupling constant.

Let us first consider the strong-coupling limit where $\beta \to 0$, as illustrated in the left panel of Fig.~\ref{fig:groundstate_O3_theta}.
Here, for concreteness, we choose a coupling of $\beta = 1/10$.
Independent of the computational approach (i.e.~using either QC or tensor network algorithms), we observe that the ground state energy density plateaus as the system size is increased.
This naively suggests that each method, individually, converges to the vacuum state and approaches a continuum limit, which is supported by the periodic boundary conditions of the system.
More importantly, we find good agreement between both QC approaches (shown in blue and orange) and the classical DMRG approach implemented via classical TNs (shown in green and red).
Again, the former is given by a simplified two-design (Q-S2D) and an MPS-inspired (Q-MPS) architecture, while the latter is labelled by the Hamiltonian truncation parameter $\lmax$.
We do not observe any significant gain in accuracy when increasing the truncation parameter.
This indicates that the system is indeed already converged to the vacuum state.
We conclude that the QC approach can capture the relevant physical properties of the ground state in the strong-coupling regime.
Indeed, this agrees with our naive expectations.
In the strong-coupling limit, the kinetic term of the Hamiltonian is dominant compared to the interaction term.
The former, however, is already diagonal in the basis at hand, thereby providing sufficient information to capture all relevant physical properties of the quantum theory.

In contrast, higher truncations of the local Hilbert spaces are necessary for the weak-coupling regime of the theory, $\beta \to \infty$.
This is illustrated in the right panel of Fig.~\ref{fig:groundstate_O3_theta}.
Here, as an example, we show the vacuum energy density at $\beta = 10$ and find that it strongly depends on the truncation parameter.
In particular, the system does not appear to be converged to the true ground state, even if the largest truncation that is computationally feasible in this scenario, $\lmax = 5/2$, is chosen\footnote{This truncation corresponds to a local Hilbert space of dimension twelve at each site of the spin chain. The computational costs for treating any Hilbert spaces beyond that increase dramatically (see also Tab.~\ref{tab:opt_num_tn}).}.
Carefully note that, therefore, the annotated error bars in the figure dramatically underestimate the total uncertainty.
Furthermore, the error bars for both approaches originate from different considerations.
As such, they have to be taken with caution when comparing both to each other.
Strikingly, neither of the Hamiltonian truncations shown here captures the weak-coupling regime of the theory sufficiently.
Therefore, our quantum simulation can only be strictly trusted in the strong-coupling limit, as it effectively amounts to a truncation characterised by $\lmax = 1/2$.
This will also play a crucial role in the discussion of the entanglement entropy of the ground state.

\subsection{The entanglement entropy of the ground state}
\label{subsec:entropy}

\begin{figure*}[t]
	\centering
	\includegraphics[width=0.75\columnwidth]{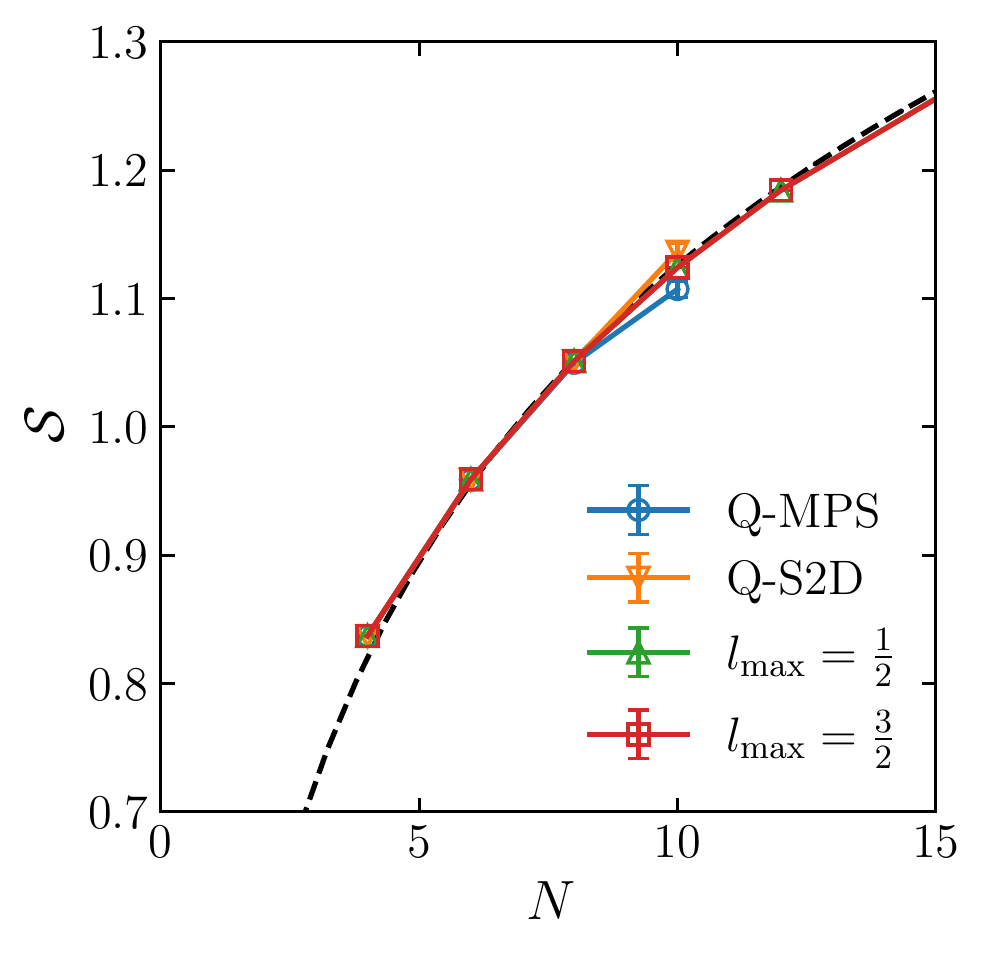}
	\hspace{2.75cm}
	\includegraphics[width=0.735\columnwidth]{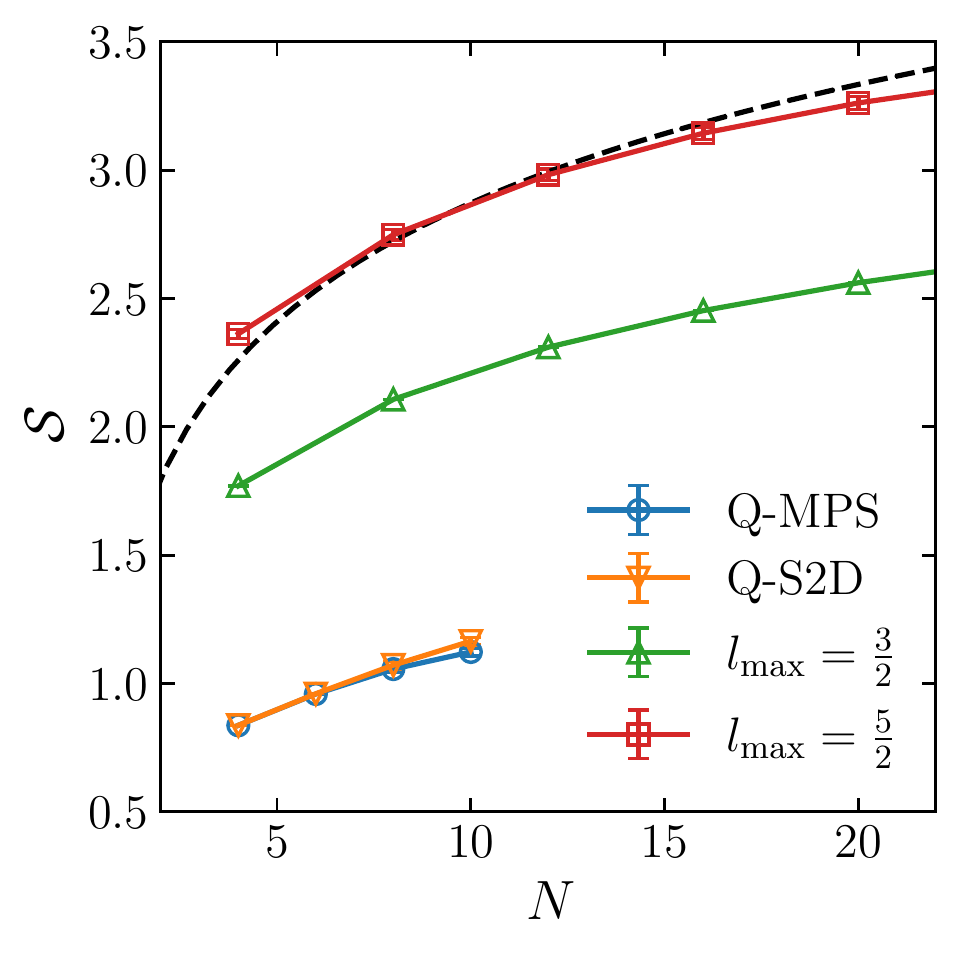}
	\caption{Half-chain entanglement entropy of the ground state as a function of the number of lattice sites in the strong-coupling (left) and weak-coupling (right) regime. (Left) We fix the coupling constant to $\beta = 1/10$. The blue and orange lines denote the simplified two-design and the MPS-inspired quantum circuits, while the green and red lines correspond to conventional TNs, characterised by the truncation parameter $\lmax$. The black-dashed line corresponds to the CFT expectation~\eqref{eq:entropyHalfChainPBC} with unit central charge, $c=1$. (Right) We fix the coupling constant to $\beta = 10$ and impose periodic boundary conditions. The colours illustrate a specific truncation of the local Hilbert space in a TN approach. The black-dashed line corresponds to the CFT expectation~\eqref{eq:entropyHalfChainPBC} with $c=2$. The error bars illustrate the uncertainties due to the optimisation algorithm (for the quantum circuits) and the DMRG truncation error cutoff (for the TNs), respectively. Carefully note that both drastically underestimate the true error done by the Hamiltonian truncation procedure (see main text).}
\label{fig:entropy_O3_theta}
\end{figure*}

An important measure to investigate the properties of a quantum many-body system is the amount of entanglement.
The entanglement entropy typically characterises the latter.
For instance, at a one-dimensional critical point where a phase transition occurs, it is well known that the entanglement entropy features a universal scaling with the system size.
Therefore, the entanglement entropy contains information on the location of the critical point and even on the central charge of the underlying CFT.
In turn, we can exploit this property to extract information on the mass gap of the quantum theory by examining the entanglement entropy of the quantum many-body state and comparing it to the underlying CFT.
We expect the gap to close in a flat space at the conformal point.\footnote{In curved space, this may be different (see, e.g.,~\cite{Hickling:2015sha}). For instance, at least naively, a quantum theory defined on a compact space introduces an intrinsic energy scale through its compactification.}

First, we define the bipartite entanglement entropy for a general quantum system by considering a subsystem of the quantum many-body state.
In this case, the von Neumann entropy is given by
\begin{equation}
	\mathcal{S} = - \Tr \left( \rho \ln \rho \right) \, ,
\end{equation}
where $\rho$ denotes the reduced density matrix of the subsystem.
Crucially, this entropy scales logarithmically with the system size in a two-dimensional CFT (at zero temperature).
For example, let us consider the entanglement entropy of the half-chain, i.e.~we choose the subsystem to be exactly half of the lattice sites. 
The half-chain entropy at a quantum critical point is then governed by the underlying CFT~\cite{Holzhey:1994we,Vidal:2002rm,Calabrese:2004eu,Calabrese:2009qy},
\begin{equation}
	\mathcal{S} = \frac{c}{3} \ln \left( \frac{N}{\pi} \right) + \mathcal{S}_0 \, .
\label{eq:entropyHalfChainPBC}
\end{equation}
Here, $c$ is the central charge of the CFT, $N$ is the (integer) system size, and $\mathcal{S}_0$ is a non-universal constant.
We have further set the lattice spacing to $a=1$.
Carefully note that this expression is only valid for a system with periodic boundary conditions.

Let us compare the behaviour of the entanglement entropy dictated by the CFT at the quantum critical point to our scenario.
As we encode each quantum many-body state directly, our approach allows us to analyse the von Neumann entropy from first principles.
In particular, we can study the half-chain entanglement entropy of the ground state constructed before.
Again, we consider the strong-coupling ($\beta \to 0$) and the weak-coupling ($\beta \to \infty$) regime separately.
Both are illustrated in Fig.~\ref{fig:entropy_O3_theta}, where we show the entanglement entropy of the vacuum state as a function of the system size.

Let us first consider the strong-coupling regime, $\beta \to 0$, shown in the left panel of Fig.~\ref{fig:entropy_O3_theta}.
Here, similar to the previous section, we choose a coupling of $\beta = 1/10$.
In this regime, we find that the half-chain entanglement entropy is in good agreement with the underlying CFT expectation given in~\eqref{eq:entropyHalfChainPBC}.
Crucially, we find the latter to be of unit central charge, $c=1$, in line with earlier results~\cite{Affleck:1987ch, Affleck:1988nt}.
This strongly indicates that the quantum theory is massless as $\beta \to 0$.\footnote{We have checked this for various other values of $\beta$ in the strong-coupling limit.}
Intriguingly, we confirm this result for each implementation of the quantum system, i.e.~for our QC approach (shown in blue and orange) and for conventional TN methods (shown in green and red).
Even for modest values of the truncation parameter, the latter appear to have converged.
This implies that, in line with the previous subsection, the system at the smallest truncation, $\lmax = 1/2$, yields a vanishing mass gap of the quantum theory.
Therefore, our approach can capture the critical behaviour of the QFT adequately.

In contrast, the situation is more complicated in the weak-coupling regime, $\beta \to \infty$.
This is shown in the right panel of Fig.~\ref{fig:entropy_O3_theta}, where we illustrate the half-chain entanglement entropy of the vacuum state at $\beta = 10$.
The Hilbert space truncation imposed within our QC approach does not sufficiently capture the weak-coupling regime of the quantum theory.
Strikingly, compared to truncations beyond $\lmax = 1/2$ (shown in green and red), the simulation is not yet converged, even as $\lmax$ is increased.
Furthermore, we would expect a central charge of $c=2$ in the weak-coupling regime due to the emergence of two massless degrees of freedom as $\beta \to \infty$.
This behaviour, however, is not correctly reproduced within our QC approach.
Therefore, the error bars shown here have to be understood as a drastic underestimation of the total uncertainty.

\begin{figure}[t]
    \centering
    \includegraphics[width=0.75\columnwidth]{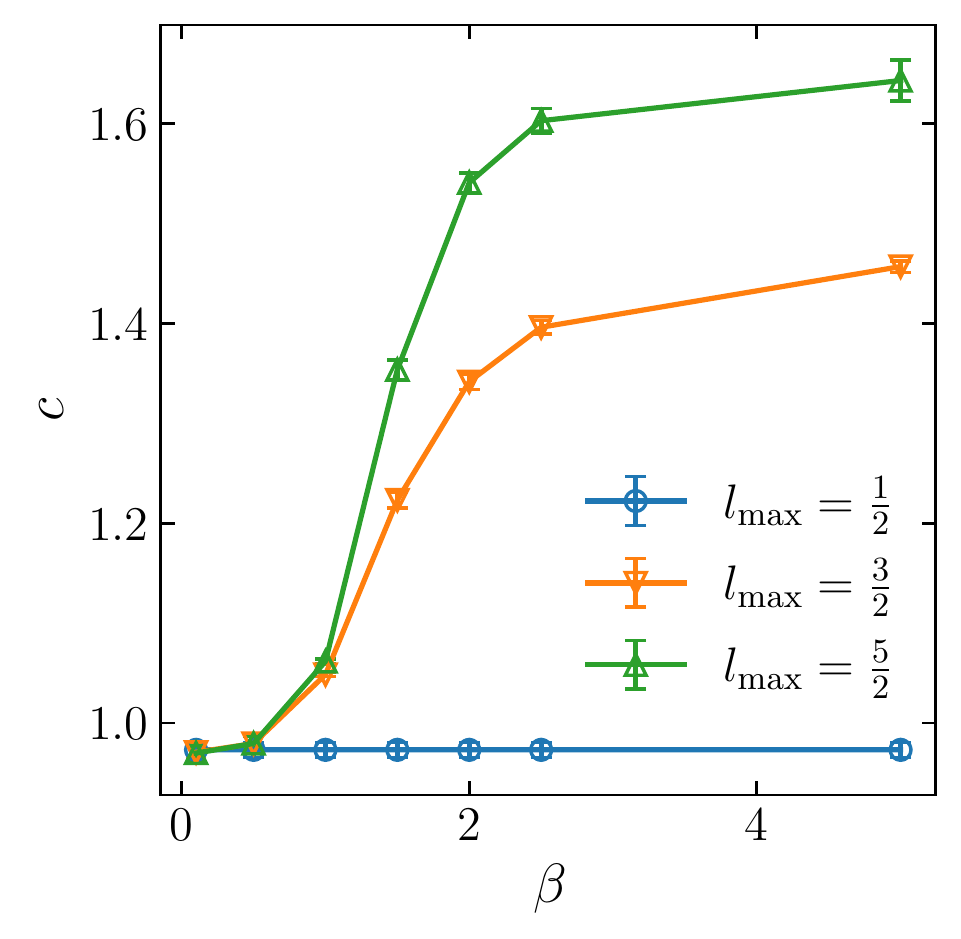}
    \caption{Central charge of the underlying CFT at the critical point as a function of the coupling constant of the quantum theory. The different colours illustrate different truncations of the local Hilbert space at each site. The error bars illustrate the uncertainties due to the DMRG truncation error cutoff. Carefully note that this drastically underestimates the true error done by the Hamiltonian truncation procedure (see main text).}
    \label{fig:ccharge}
\end{figure}

This limitation is further illustrated in Fig.~\ref{fig:ccharge}.
Here, we show the central charge of the underlying CFT (according to the logarithmic scaling~\eqref{eq:entropyHalfChainPBC}) as a function of the coupling $\beta$ for different truncations of the local Hilbert space at each site.
At first sight, within our QC approach, which effectively amounts to $\lmax = 1/2$, the central charge naively appears to be independent of $\beta$.
Clearly, this cannot be correct, as we would expect a smooth, monotonic interpolation between $c=1$ in the strong-coupling limit, and $c=2$ in the weak-coupling limit, according to the $c$-theorem~\cite{Zamolodchikov:1986gt}.
Higher values of the truncation parameter beyond $\lmax = 1/2$, which we implement via conventional TN methods, appear to perform better as $\beta \to \infty$.
However, at present, any QC architecture embedding a truncation beyond $\lmax = 1/2$ is computationally not feasible (we will comment on this issue in the following subsection).
Therefore, TN methods prove to be more reliable in the weak-coupling regime of the nonlinear sigma model at $\theta=\pi$.
In summary, $\lmax$ must be sufficiently large to capture the quantum system's critical behaviour in the weak-coupling regime.
This is in line with recent observations~\cite{Tang:2021uge}.

\subsection{Discussion}
\label{subsec:discussion}

Despite the success of the VQE approach, there is an explicit limitation to the ability to perform genuinely quantum computations.
Due to the infinite-dimensional local Hilbert spaces associated with the nonlinear sigma model (or any other QFT), a Hamiltonian truncation is necessary regardless of the computational scheme.
In this work, we use a truncation scheme that effectively reduces the local Hilbert space at each site to two dimensions.
While it allows us to simulate a sizable number of lattice sites within our QC approach, it eliminates most of the information inherent to the Hamiltonian of the theory.
For example, using a truncation parameter of $\lmax = 1/2$ enables us to simulate up to ten spin chain sites without losing the optimisation algorithm's sensitivity.
Although it may, in principle, be possible to simulate the reduced density matrix of a quantum circuit for up to 20 lattice sites, we observe that our optimisation algorithm's performance significantly reduces beyond ten lattice sites.

In practice, the quantum simulation we employ in this work requires each local Hilbert space to be of dimension $2^d$.
However, this cannot be achieved with any naturally available truncation at half-integer $\lmax$.
We would therefore need to, admittedly somewhat artificially, add another truncation by, for example, considering another singular value decomposition of the Hamiltonian \emph{after} truncating it in the usual way.
We leave this for future work.
However, given that in the presented approach, we can effectively simulate up to ten lattice sites, i.e.~ten qubits, we are doubtful that any more extensive truncation is feasible to optimise on a quantum device due to the barren plateaus.
Given the limitations of the gradient descent method, our approach is currently limited to a specific coupling regime of the quantum theory (cf.~also Fig.~\ref{fig:ccharge}).

Nevertheless, despite the limitations originating from the optimisation algorithm, we observe that the ground state of the quantum theory can be represented more efficiently by a QC approach compared to classical TN methods.
For instance, Tab.~\ref{tab:opt_num} shows the number of layers and trainable parameters required to achieve the presented results for a simplified two-design and an MPS-inspired quantum circuit for a truncation parameter of $\lmax = 1/2$.
For comparison, in Tab.~\ref{tab:opt_num_tn}, we present the maximum bond dimensions and the number of parameters required to achieve the same accuracy within the conventional DMRG approach implemented via classical TNs.
We illustrate the truncation parameters $\lmax = 1/2$ and $\lmax = 3/2$, respectively.
Even in the simplest four-site example, the classical TN approach requires almost twice as many parameters as the QC approach.

\begin{figure}[t]
    \centering
    \includegraphics[width=0.75\columnwidth]{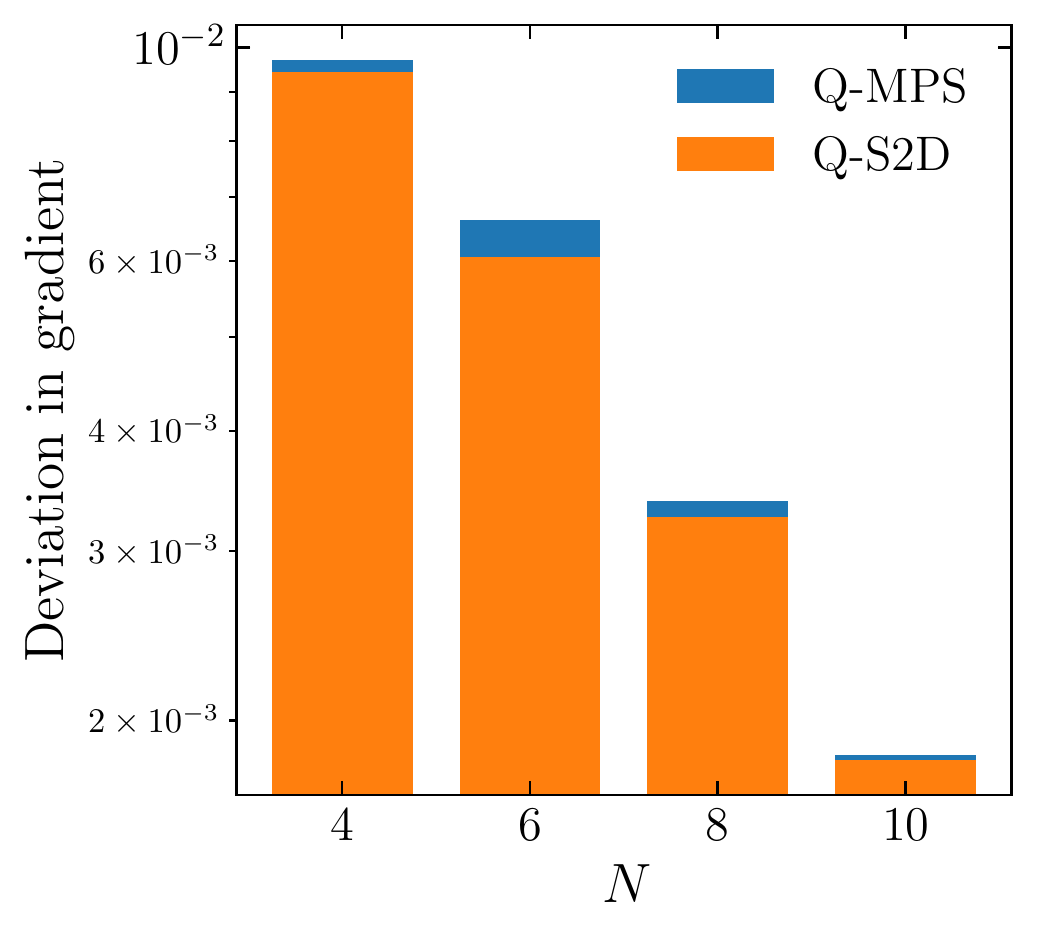}
    \caption{Mean standard deviation in the gradient of the trainable parameters with respect to the expectation value. Orange shows the circuit with the simplified two-design and blue shows the circuit with MPS architecture.}
    \label{fig:std_grad}
\end{figure}

\begin{figure}[t]
    \centering
    \includegraphics[scale=0.42]{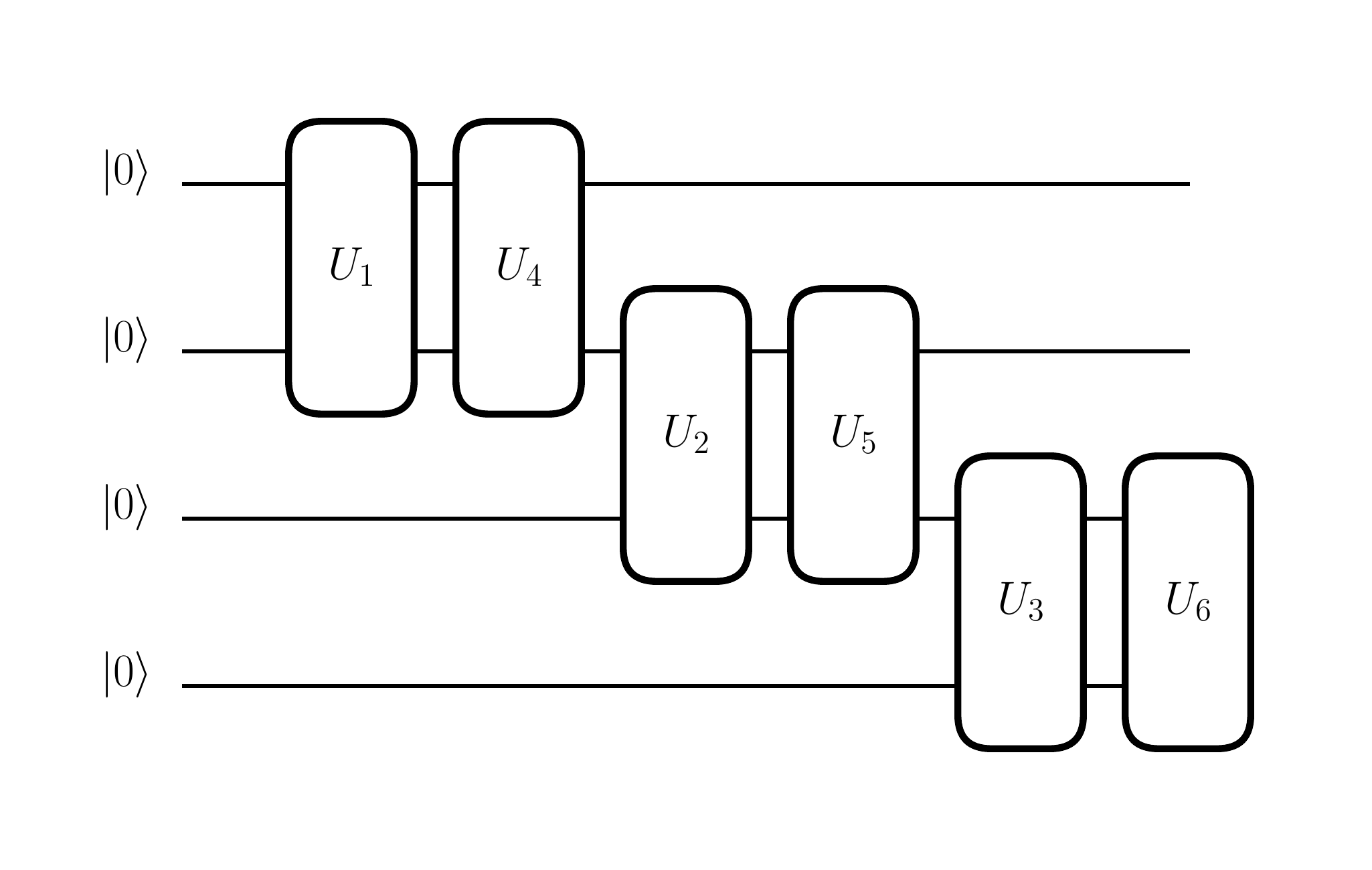}
    \caption{Quantum circuit similar to Fig.~\ref{fig:vqe} but as a block-extended MPS-inspired architecture.}
    \label{fig:qmps_block}
\end{figure}

In practice, as mentioned above, with an increasing number of sites, gradient methods are increasingly challenging to optimise our VQE approach.
For instance, Fig.~\ref{fig:std_grad} shows the standard deviation in the value of the gradient for 100 independent initialisations of the quantum circuit for the simplified two-design and the MPS-inspired architecture.
It is clear that, even with unlimited computing resources, gradient-based methods are insufficient to optimise any given quantum circuit for the QFT problem at hand for a large number of sites.
We have also tested our VQE ansatz with gradient-free optimisation algorithms, such as \textsc{COBYLA}~\cite{Powell2007AVO}.
However, converging to a minimum requires significantly more computation time.
In our tests, we have not been able to achieve the same precision as shown in Fig.~\ref{fig:groundstate_O3_theta} and~\ref{fig:entropy_O3_theta} for a ten-site scenario within 100,000 iterations.\footnote{It is, however, possible to thoroughly investigate four and six lattice sites within a reasonable computing time.}
Overall, this significantly limits the applicability of the VQE approach.
Therefore, it is essential to investigate improved optimisation algorithms to simulate such QFT problems on a larger scale.\footnote{For instance, a detailed inspection of barren plateaus in TN-inspired quantum circuits can be found in~\cite{Martin:2022evx}.}
For instance, it has been shown that quantum annealing setups can improve upon the classical gradient-based methods~\cite{Abel:2022lqr, Abel:2021fpn}.
These can be applied to universal superconducting circuits via digitised adiabatic quantum computing techniques~\cite{Barends:2016, CoelloPerez:2021jkh}.
A recent study has also shown that \emph{perturbative gadgets} can be used to improve the optimisation landscape for many-body Hamiltonian simulations~\cite{Cichy:2022nzq}.

In addition, we have investigated various architectures for the quantum circuits at hand.
Despite having the same trainable parameters, we observe that specific architectures cannot correctly capture the entanglement entropy information shown in Fig.~\ref{fig:entropy_O3_theta}.
For instance, we have tested an MPS-inspired architecture augmented with a block-extended version, as shown in Fig.~\ref{fig:qmps_block}.
Although this architecture involves the same number of trainable parameters and a similar general structure as the right panel of Fig.~\ref{fig:vqe}, it cannot find the ground state with the same accuracy.
While the latter deviation is not dramatic in its own right, the entanglement entropy is highly sensitive to this procedure.
Therefore, while this architecture may, in principle, have the potential to represent the underlying quantum system, it cannot adequately capture the entanglement properties (and hence the critical behaviour) of the quantum theory.

It is essential to note that this work has been limited to a quantum simulation implementation without shot noise.
Although we observe that for a small number of sites it is possible to recover the physical properties of the underlying theory, shot noise becomes overwhelmingly large.
This is because an increase in the number of lattice sites requires deeper circuits.
This indicates that a more realistic analysis of nonlinear sigma models on a quantum computer will require the implementation of error mitigation techniques which are beyond the scope of this paper.
Hence, strictly speaking, this demonstration is only for the purpose of validating the correctness of the variational quantum algorithm.

\section{Conclusion}
\label{sec:conclusion}

Digital quantum algorithms constitute a promising approach to the numerical investigation of nontrivial QFT dynamics.
As proof of principle, in this work, we have studied the mass gap of a two-dimensional $O(3)$ nonlinear sigma model augmented with a topological $\theta$-term.
For simplicity, we have chosen $\theta = \pi$.
Using digital quantum methods, we confirm that the quantum theory is massless by comparing the entanglement entropy of the vacuum state to the underlying CFT at the critical point.

Crucially, our approach relies on a Hamiltonian truncation.
First, we discretise the quantum theory on a lattice to apply a suitable truncation of the local Hilbert space at each lattice site.
At the same time, this procedure profoundly restricts the applicability of digital quantum algorithms to realistic QFTs.
Here, we have pointed out the possibilities and, more importantly, the limitations of these algorithms in a simple yet nontrivial setup.
Our results indicate that, even with ideal quantum devices, it will be challenging to use classical gradient-based optimisation techniques to correctly reproduce the physical properties of the nonlinear sigma model augmented with a topological $\theta$-term.
While the strong-coupling regime, where $\beta \to 0$, is in line with theoretical expectations from the underlying CFT, the physical properties of the quantum theory in the weak-coupling limit are not appropriately captured.
The latter is mainly due to the strict Hamiltonian truncation necessary to simulate the theory on a quantum device in the first place.
We find that, in this scenario, conventional TN methods, such as DMRG, outperform the QC approach in accuracy.
Given that in our setup, we can simulate up to ten lattice sites, i.e.~ten qubits, effectively.
It would be interesting to explore alternative Hamiltonian truncation schemes of the quantum theory in the future (see, e.g.,~\cite{Alexandru:2019ozf, Alexandru:2022son}).

Similarly, our results indicate that, even with ideal quantum devices, classical optimisation techniques pose a fundamental challenge to the numerical investigation of realistic nonlinear sigma models. This is mainly due to the barren plateau problem that has been observed in wider circuits.
Perhaps the use of fully quantum optimisation techniques, such as quantum annealing, may be essential to investigate realistic QFTs with noise-tolerant quantum devices in the future.

Beyond the technical challenges ahead, it is similarly not straightforward to find a Hamiltonian description of a realistic QFT suitable for investigation on quantum devices.
For instance, it would be interesting to generalise the results presented in this work to two-dimensional nonlinear sigma models with a topological term at arbitrary values of $\theta$.
In summary, we conclude that systematically constructing arbitrary Hamiltonian truncation schemes of (semi)realistic QFTs merits further exploration.

\begin{acknowledgments}
SS is funded by the Deutsche Forschungsgemeinschaft (DFG, German Research Foundation) -- 444759442.
\end{acknowledgments}

\appendix

\section{Operator matrix elements}
\label{app:matrix_elements}

To find the (low-lying) spectrum of the Hamiltonian $H$, it is necessary to know how the operators belonging to $H$ act on the local Hilbert space at each site.
As we have written our theory in terms of angular momentum operators, we can use their well-known algebra, $\left[L^a, L^b\right] = i \epsilon_{abc} L^c$, to obtain their corresponding matrix elements.
For instance, the ladder operators of the algebra,
\begin{equation}
    L^{\pm} = L^x \pm i L^y \, ,
\end{equation}
act (up to a complex phase) on the Hilbert space of the $k$-th site as (see, e.g.,~\cite{Wu:1976ge})
\begin{equation}
    L_k^{\pm} \ket{qlm}_k = \sqrt{\left(l \mp m\right)\left(l \pm m + 1 \right)} \ket{q,l,m \pm 1}_k \, .
\end{equation}
From this, it is evident that the first term, corresponding to the rotational energy of each site, is already diagonal.
Therefore, in the case of the $O(3)$ nonlinear sigma model, the matrix elements of the operator $\vec{L}^2$ are straightforwardly given by eq.~\eqref{eq:O3_ladder}.

The second term, however, still needs to be diagonalised.
To do this, we first define the complex fields (analogous to the ladder operators)
\begin{equation}
	n^{\pm} = \frac{1}{\sqrt{2}} \left(n^x \pm i n^y\right) \, ,
\end{equation}
such that the potential term for each pair of neighbouring sites can be written as
\begin{equation}
	\vec{n}_k \vec{n}_{k+1} = n_k^+ n_{k+1}^- + n_k^- n_{k+1}^+ + n_k^z n_{k+1}^z \, .
\end{equation}
The matrix elements of each operator $\vec{n}_k$ can then be obtained by realizing that, in the position-space representation, they are closely related to the so-called monopole harmonics~\cite{Wu:1976ge, Wu:1977qk}.
Therefore, the matrix elements will be proportional to Wigner-$3j$ symbols~\cite{Wu:1977qk} (see also~\cite{Tang:2021uge}),
\begin{equation}
\begin{split}
	\Braket{ql^{\prime} m^{\prime} | X_{M} | qlm} =& (-1)^{q+m^{\prime}+l+l^{\prime}+1} \sqrt{\left(2l^{\prime} + 1\right) \left(2l + 1\right)} \\
		&\times \begin{pmatrix}
				l^{\prime} & 1 & l \\
				-q & 0 & q
				\end{pmatrix}
				\begin{pmatrix}
				l^{\prime} & 1 & l \\
				-m^{\prime} & M & m
				\end{pmatrix} \, ,
\end{split}
\end{equation}
where we can finally identify $n^{\pm} = \mp X_{\pm 1}$ and $n^z = X_0$.

\section{Embedding the Hamiltonian into a Quantum Circuit}
\label{app:hamiltonian_embedding}

In practice, in order to embed a given Hamiltonian into a quantum circuit, it has to be written in terms of Pauli matrices.
Since we have already obtained the operator matrix elements associated with the Hamiltonian in an arbitrary truncation in Appendix~\ref{app:matrix_elements}, we can easily determine their representation by Pauli matrices.
They are given by
\begin{equation}
	H_0 = \frac{3}{4} I \, , \, n^\pm = -\frac{1}{3\sqrt{2}} \left(\sigma_X \pm i\sigma_Y\right) \, , \, n^z = \frac{1}{3} \sigma_Z \, ,
\end{equation}
where $I$ represents the $2 \times 2$ identity matrix and $\sigma_i$ denote the well-known Pauli matrices.
Furthermore, $H_0$ represents the kinetic term of the Hamiltonian.

\bibliography{references,references_noninspire}

\end{document}